\documentclass[conference]{IEEEtran}
\IEEEoverridecommandlockouts
\usepackage{cite}
\usepackage{amsmath,amssymb,amsfonts}
\usepackage{algorithmic}
\usepackage{graphicx}
\usepackage{textcomp}
\usepackage[]{authblk}
\usepackage{xcolor}
\usepackage{comment}
\def\BibTeX{{\rm B\kern-.05em{\sc i\kern-.025em b}\kern-.08em
    T\kern-.1667em\lower.7ex\hbox{E}\kern-.125emX}}

\usepackage{graphicx}	
\usepackage{amsmath}	
\usepackage{comment}
\usepackage{xspace} 
\usepackage{mathtools}
\usepackage{hyperref}
\usepackage{float}
\usepackage{bm}
\usepackage{pdflscape}
\usepackage{pdfpages}
\usepackage{stmaryrd}
\usepackage{setspace}
\DeclareFontFamily{U}{mathx}{}
\DeclareFontShape{U}{mathx}{m}{n}{<-> mathx10}{}
\DeclareSymbolFont{mathx}{U}{mathx}{m}{n}
\DeclareMathAccent{\widecheck}{0}{mathx}{"71}

\usepackage{multirow} \usepackage{threeparttable}
\usepackage{booktabs} \usepackage{stfloats}

\newcommand{\PACO}{{\texttt{PACO}}\xspace} 
\newcommand{\PACOII}{{\texttt{PACO ASDI}}\xspace} 
\newcommand{\dPACO}{{\texttt{deep PACO}}\xspace} 
\newcommand{\dPACOII}{{\texttt{deep PACO ASDI}}\xspace} 

\newcommand*{\V}[1]{\boldsymbol{#1}}   
\newcommand*{\M}[1]{\mathbf{#1}}       
\newcommand*{\TransposeLetter}{\hspace*{-.25ex}\top\hspace*{-.25ex}}
\newcommand*{\T}{^{\TransposeLetter}} 
\DeclareFontFamily{U}{mathx}{\hyphenchar\font45}
\DeclareFontShape{U}{mathx}{m}{n}{<-> mathx10}{}
\DeclareSymbolFont{mathx}{U}{mathx}{m}{n}
\DeclareMathAccent{\widebar}{0}{mathx}{"73}

\DeclarePairedDelimiterX{\Paren}[1]{(}{)}{#1}
\DeclarePairedDelimiterX{\Brace}[1]{\{}{\}}{#1}
\DeclarePairedDelimiterX{\Brack}[1]{[}{]}{#1}
\DeclarePairedDelimiterX{\Abs}[1]{\rvert}{\lvert}{#1}
\DeclarePairedDelimiterX{\Norm}[1]{\lVert}{\rVert}{#1}
\DeclarePairedDelimiterX{\Avg}[1]{\langle}{\rangle}{#1}
\DeclarePairedDelimiterX{\Round}[1]{\lfloor}{\rceil}{#1}
\DeclarePairedDelimiterX{\Floor}[1]{\lfloor}{\rfloor}{#1}
\DeclarePairedDelimiterX{\Ceil}[1]{\lceil}{\rceil}{#1}
\DeclarePairedDelimiterX{\Inner}[2]{\langle}{\rangle}{#1,#2}

\DeclareMathOperator{\Trace}{tr}

\DeclarePairedDelimiterXPP{\Expect}[1]{\mathbb{E}}(){}{#1}

\newcommand*{\estim}[1]{\widehat{#1}}
\newcommand{\Sest}{\widehat{\M{S}}}

\usepackage[squaren,Gray]{SIunits}

\usepackage{numprint}

\makeatletter
\def\widebreve{\mathpalette\wide@breve}
\def\wide@breve#1#2{\sbox\z@{$#1#2$}%
     \mathop{\vbox{\m@th\ialign{##\crcr
\kern0.08em\brevefill#1{0.8\wd\z@}\crcr\noalign{\nointerlineskip}%
                    $\hss#1#2\hss$\crcr}}}\limits}
\def\brevefill#1#2{$\m@th\sbox\tw@{$#1($}%
  \hss\resizebox{#2}{\wd\tw@}{\rotatebox[origin=c]{90}{\upshape(}}\hss$}
\makeatletter

\begin{document}

\title{
Combining multi-spectral data with statistical and deep-learning models for improved exoplanet detection in direct imaging at high contrast
{}
\thanks{The work of OF, ML, and AML was supported in part by the European Research Council (ERC) under the \textit{European Union's Horizon 2020 research and innovation program} (COBREX; n° 885593). The work of TB and JP was supported in part by the Inria/NYU collaboration, the Louis Vuitton/ENS chair on artificial intelligence and the French Agence Nationale de la Recherche (ANR) as part of the \textit{Investissements d'avenir program} (PRAIRIE 3IA Institute; n° 19-P3IA-0001). The work of JM was supported in part by the ERC (SOLARIS; n° 714381) and by the ANR (3IA MIAI, n° 19-P3IA-0003).}
}

\author{Olivier Flasseur$^1$}
\author{Théo Bodrito$^2$}
\author{Julien Mairal$^3$}
\author{Jean Ponce$^{2,4}$}
\author{Maud Langlois$^1$}
\author{Anne-Marie Lagrange$^{5,6}$}
\affil{\small $^1$Centre de Recherche Astrophysique de Lyon, CNRS, Univ. de Lyon, Univ. Claude Bernard Lyon 1, ENS de Lyon, France\protect\\ $^2$Département d'Informatique de l'{\'E}cole Normale Supérieure (ENS-PSL, CNRS, Inria), France\protect\\ $^3$Univ. Grenoble Alpes, Inria, CNRS, Grenoble INP, LJK,  France\protect\\ $^4$Courant Institute of Mathematical Sciences, Center for Data Science, New York Univ., USA\protect\\ $^5$Laboratoire d'{\'E}tudes Spatiales et d'Instrumentation en Astrophysique, Obs. de Paris, Univ. PSL, Sorbonne Univ., Univ. Diderot, France\protect\\ $^6$Univ. Grenoble Alpes, Institut de Planétologie et d'Astrophysique de Grenoble, France}

\maketitle

\begin{abstract}
Exoplanet detection by direct imaging is a difficult task: the faint signals from the objects of interest are buried under a spatially structured nuisance component induced by the host star. The exoplanet signals can only be identified when combining several observations with dedicated detection algorithms. 
In contrast to most of existing methods, we propose to learn a model of the spatial, temporal and spectral characteristics of the nuisance, directly from the observations. In a pre-processing step, a statistical model of their correlations is built locally, and the data are centered and whitened to improve both their stationarity and signal-to-noise ratio (SNR). A convolutional neural network (CNN) is then trained in a supervised fashion to detect the residual signature of synthetic sources in the pre-processed images. 
Our method leads to a better trade-off between precision and recall than standard approaches in the field. It also outperforms a state-of-the-art algorithm based solely on a statistical framework. Besides, the exploitation of the spectral diversity improves the performance compared to a similar model built solely from spatio-temporal data.
\end{abstract}

\begin{IEEEkeywords}
detection, supervised deep learning, matched filter, multi-variate data, correlated data
\end{IEEEkeywords}

\section{Introduction}
\label{sec:introduction}
 
Alongside indirect methods for detecting exoplanets (e.g., based on the monitoring of transits or of the radial velocities), direct imaging \cite{traub2010direct} from Earth is a method of choice to detect and  characterize the physical properties (e.g., orbit, atmospheric composition) of exoplanets. This characterization is carried out by comparing the estimated astrometry and spectral energy distribution (SED) with physical models. However, the low number of exoplanets detected and characterized with this technique (a dozen over the last twelve years) testifies to the difficulty of this long-term goal. This is explained by the very large difference in \textit{contrast}, typically greater than five orders of magnitude in the infrared, between the exoplanets and their host star. Beyond extreme adaptive optics compensating wavefront distortions and the coronagraph partly masking out the star, which are implemented in cutting-edge observational facilities, the detection of faint sources requires a dedicated processing of the data.

We recall in Section \ref{sec:principle} the key principles of high-contrast imaging (HCI) and of the state-of-the-art processing methods. We describe in Section \ref{sec:proposed_algorithm} the proposed method, which is an extension to multi-spectral data of our very recent work dedicated to the processing of spatio-temporal observations \cite{dPACO}. We compare in Section \ref{sec:results} the detection performance of the proposed method to state-of-the-art algorithms of the field on data from the SPHERE instrument currently operating at the Very Large Telescope (VLT, Chile).

\section{Principle of Direct Imaging at High Contrast}
\label{sec:principle}

\begin{figure}
	\begin{center}
		\includegraphics[width=0.42\textwidth]{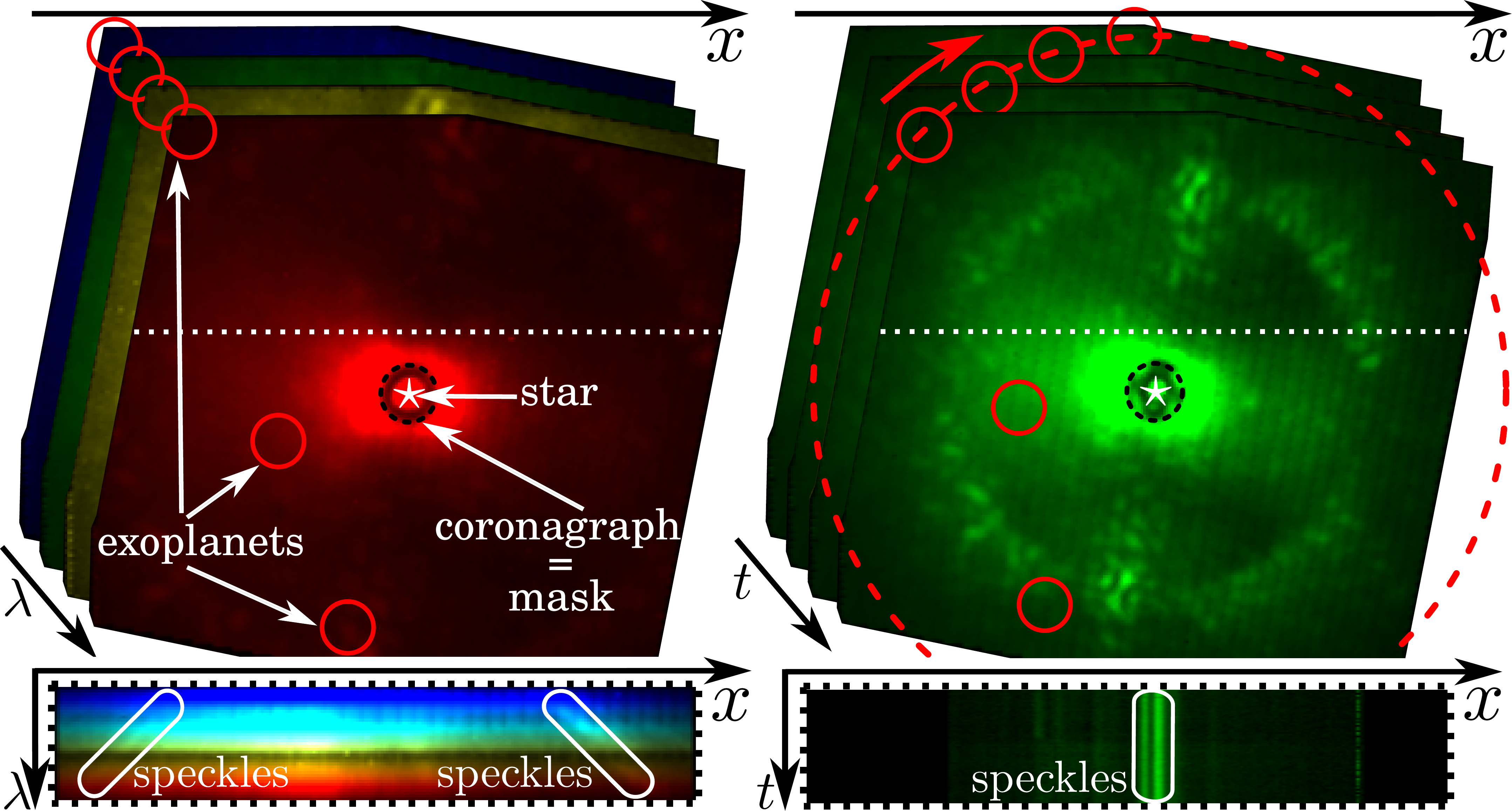}
		\caption{Illustration of a dataset form the SPHERE-IFS instrument. Left: images at different wavelengths. Right: images at different times (at $\lambda_{25} = 1.4\,\micro\meter$). Red circles represent the locations of three known exoplanets whose signatures are too faint to be detected without additional processing. Bottom: spatio-spectral and spatio-temporal slice cuts along the white dashed line.}
	\label{fig:ifs_data}
	\end{center}
\end{figure}

A classical observation technique in HCI is angular differential imaging (ADI) that consists in recording a 3-D dataset (2-D + time) over a series of short exposures during a few hours of observation. During the acquisition, all the off-axis point-like sources follow an apparent circular motion (typically, about a few tens of degrees) around the optical axis due to the Earth's rotation. Since the pupil of the telescope is kept fixed, the other structures of the background remain temporally quasi-static. Due to the diffraction and to residual optical aberrations, stellar leakages take the form of spatially correlated speckles. Speckles and other additive sources of noise (thermal background, detector readout, photon noise) form a \textit{nuisance component} that strongly contaminates the images formed downstream of the coronagraph. Besides, there is a significant variability in the amplitude and in the structure of the spatial correlations of the nuisance, which is stronger near the star than farther away. Spectral differential imaging (SDI) is a complementary observation technique allowing to capture images in several spectral channels simultaneously with an integral field spectrograph (IFS). With SDI, the speckle pattern scales approximately linearly with the wavelength due to the  diffraction. The temporal and spectral diversity induced by ADI and SDI can be combined advantageously. This hybrid technique, called angular and spectral differential imaging (ASDI), produces 4-D datasets (2-D + time + spectral). 
In this paper, we consider ASDI. As an example, the SPHERE-IFS instrument produces series of $T \simeq 100$ temporal exposures with $N = 290 \times 290$ pixels each in $L=39$ spectral bands (between $\lambda_1 = 0.9 \, \micro\meter$ and $\lambda_{39} = 1.6 \, \micro\meter$), see Fig. \ref{fig:ifs_data}.  

To unmix the exoplanet signals from the nuisance, state-of-the-art methods combine the different temporal and spectral images \cite{pueyo2018direct}. A simple solution consists in subtracting the temporal mean to each frame of a given spectral channel to attenuate speckles. The residual images are then rotated and scaled to compensate for the (predictable) motion of the off-axis sources, so that their signals are spatially co-localized. The resulting stack is finally combined, e.g., by taking the median, as in the cADI method. Other methods resort to a principal component analysis (PCA) to build an empirical model of the nuisance. These two baseline algorithms \cite{pueyo2018direct} are used routinely in HCI. 
Approaches based on detection theory are more efficient because they model explicitly the attenuation of the sough objects induced by image subtraction. In particular, the \PACO algorithm \cite{PACO} and its extension \PACOII \cite{PACOII} to multi-spectral data build a statistical model of the nuisance by capturing their spatial correlations and fluctuations at the scale of small patches of a few tens of patches. 
However, it remains room for improvement due to the approximate fidelity of this statistical model to the observations. In this context, we proposed very recently \dPACO \cite{dPACO}, a data-driven algorithm for ADI observations combining the statistical model of \PACO with supervised deep learning. \dPACO performs better than existing methods, including approaches based on deep learning that remain limited by a high false alarm rate. 
In this paper, we present \dPACOII, extending \dPACO to multi-spectral data. 

\section{Proposed Algorithm}
\label{sec:proposed_algorithm}

\subsection{Direct Model of the Observed Intensity}
\label{subsec:direct_model}

An ASDI dataset $\V r$ in $\mathbb{R}^{N \times T \times L}$ is formed by $N$-pixels images, recorded at different times $t$ in $\llbracket 1 ; T \rrbracket$ and in different spectral channels $\ell$ in $\llbracket 1 ; L \rrbracket$. The contribution $\V r_{\ell}$ in $\mathbb{R}^{N \times T}$ is: 
\begin{equation}
	\V r_\ell = \V f_\ell + \sum\limits_{p=1}^P \alpha_{p,\ell} \, \V h_\ell (\phi_p)\,,
	\label{eq:image_formation_model}
\end{equation}
where $\V f_\ell$ in $\mathbb{R}^{N \times T}$ and $\V h_\ell$ in $\mathbb{R}^{N \times T}$ are respectively the contribution of the nuisance and of any point-like source taking the form of the off-axis point-spread function (PSF), at spectral channel $\ell$. The contribution of a source $p \in \llbracket 1 ; P \rrbracket$ is centered at location $\mathcal{F}_{t,\ell}(\phi_p)$ in the $t$-th image of the $\ell$-th channel, where $\phi_p$ is its initial location on an image at a time $t_{\text{ref}}$ and spectral channel $\lambda_{\text{ref}}$ of reference. $\mathcal{F}_{t,\ell}$ is a deterministic geometrical transform (a composition of a rotation and of a radial translation with respect to the star) to account for the apparent rotation induced by ADI and to the translation induced by spatial scaling of factor $\lambda_{\text{ref}}/\lambda_\ell$ applied beforehand to spatially co-align the nuisance's  structures, see Fig. \ref{fig:ifs_data}.

\subsection{Step 1: Pre-processing by Statistical Modeling}
\label{subsec:pre_processing}

Based on our previous work \cite{PACOII}, we model the random fluctuations of the nuisance component $\V f$ by a statistical model whose parameters are self-calibrated on the observations. The model is local, i.e., it is built at the scale of patches of a few tens of pixels to account for the high non-stationarity of the nuisance component. We consider non-overlapping square patches of $K$ pixels, and we note $\mathbb{P}$ the set of locations paving the whole field of view. We model the distribution of a patch $\V f_{n, t, \ell}$ centered at pixel $n$, at time $t$, and channel $\ell$ by a scaled multi-variate Gaussian: $\mathcal{N}(\V m_{n,\ell}, \sigma_{n,t,\ell}^2\M C_n)$. The sample estimators $\lbrace \widehat{\V m}_{n,\ell}\,, \widehat{\sigma}_{n,t,\ell}^{2} \,, \widehat{\M S}_n\rbrace$ of $\lbrace \V m_{n,\ell}\,, \sigma_{n,t,\ell}^{2} \,, \M C_n\rbrace$ are obtained from the $T\,L$ patches $\V r_n$ in $\mathbb{R}^{K \times T \times L}$ in the maximum-likelihood sense with a fixed-point iterative scheme:
\begin{small}
\begin{equation}
	\begin{cases}
		\widehat{\V m}_{n,\ell} =  \big(\sum\limits_t \widehat{\sigma}_{n,t,\ell}^{-2}\big)^{-1} \sum\limits_{t} \widehat{\sigma}_{n,t,\ell}^{-2} \, \V r_{n,t,\ell}\,,\\
		\widehat{\M S}_n = \frac{1}{T\,L} \sum\limits_{t, \ell}  \widehat{\sigma}_{n,t,\ell}^{-2} (\V r_{n,t,\ell} - \widehat{\V m}_{n,\ell}) (\V r_{n,t,\ell} - \widehat{\V m}_{n,\ell})\T\,,\\
		\widehat{\sigma}_{n,t,\ell}^{2} = K^{-1} (\V r_{n,t,\ell} - \widehat{\V m}_{n,\ell})\T \, \widehat{\M S}_n^{-1} (\V r_{n,t,\ell} - \widehat{\V m}_{n,\ell})\,,
	\end{cases}
	\label{eq:sample_estimators}
\end{equation}
\end{small}
\noindent where the scaling factors $\lbrace \widehat{\sigma}_{n,t,\ell}^2 \rbrace_{t=1:T, \ell=1:L}$ are the empirical variances of the (whitened) residuals, that allow to neutralize patches with outliers. The total number of samples involved in the computation of $\widehat{\M S}_n$ being lower than the number $K$ of pixels in a patch, the sample covariance $\widehat{\M S}_n$ is very noisy and can be rank deficient. As in our previous work \cite{PACOII}, we regularize it by \textit{shrinkage} \cite{chen2010shrinkage}, i.e. the resulting estimator is a convex combination between the low bias/high variance estimator $\widehat{\M S}_n$ and a high bias/low variance estimator $\widehat{\M F}_n$: 
\begin{equation}
	\estim{\M C}_n = (1 - \widehat{\rho}_n)\,\widehat{\M S}_n + \widehat{\rho}_n\,\widehat{\M F}_n\,,
	\label{eq:shrunk_covariances}
\end{equation}
where $\widehat{\M F}_n$ is a diagonal matrix with only the sample variances, and the hyper-parameter $\widehat{\rho}_n$ setting a bias-variance trade-off. The latter is estimated \cite{dPACO} for each patch location $n$ through:
\begin{equation}
  \estim\rho_n= \frac{
    \Trace\Paren[\big]{\Sest_{n}^2}
    + \Trace^2\Paren[\big]{\Sest_{n}}
    - 2\sum_{k=1}^K\Brack[\big]{\Sest_{n}}_{kk}^2
  }{
    (Q+1)\,\Paren[\Big]{
      \Trace\Paren[\big]{\Sest_{n}^2}
      - \sum_{k=1}^K\Brack[\big]{\Sest_{n}}_{kk}^2
    }
  } \,,
   \label{eq:shrinkfactor}
\end{equation}
with $Q = \big( \sum_{t,\ell} \widehat{\sigma}_{n,t,\ell}^{-2} \big)^2 / \big( \sum_{t,\ell} \widehat{\sigma}_{n,t,\ell}^{-4} \big)$ the equivalent number of patches involve in the computation of $\widehat{\M C}_n$ in the presence of the scaling factors $\lbrace \widehat{\sigma}_{n,t,\ell}^2 \rbrace_{t=1:T, \ell=1:L}$. Given the statistics of the nuisance, the pre-processed images $\widetilde{\V r}$ in $\mathbb{R}^{N \times T \times L}$ (i.e., after centering and whitening) are obtained by:
\begin{equation}
	\widetilde{\V r}_{n,t,\ell} = \M W_{n,t,\ell} \, \V r_{n,t,\ell} = \widehat{\sigma}_{n,t,\ell}^{-1} \, \widehat{\M L}_{n}\T \, \left( \V r_{n,t,\ell} - \widehat{\V m}_{n,\ell} \right) \,,
	\label{eq:whitening}
\end{equation}
with $\widehat{\M L}_n$ the Cholesky's factorization of $\widehat{\M C}_n^{-1}$ (i.e., $\widehat{\M L}_n \widehat{\M L}_n\T = \widehat{\M C}_n^{-1}$). Most of the spatial structures of $\V r$ are removed in $\widetilde{\V r}$.

\subsection{Step 2: Exoplanet Detection by Supervised Deep Learning}
\label{subsec:deep_learning}

\subsubsection{Training basis}
\label{subsubsec:training_basis}

Starting from a temporo-spectral series of pre-processed images, we aim to infer a detection map $\widehat{\V y}$ in $\left[0 ; 1 \right]^{M}$, where each pixel-value represents the \textit{pseudo-probability} that a source is centered at that location at time $t_\text{ref}$. Obtaining real ground truth data is challenging in HCI: a few real sources have been detected and undiscovered sources might still be present in the data. We thus opt for a synthetic training strategy: the training set consists of $S$ pairs $\lbrace \widebreve{\V r}^{[s]} ; \V y^{[s]} \rbrace_{s=1:S}$ of samples/ground truths, resulting from the massive injection of point-like sources. Besides, we train a different model for each ASDI dataset due to the high variability of the nuisance component from one observation to the other. This data-dependence prevents generalizing the learned detector to a new set of pre-processed data. These peculiarities imply the design of a data-augmentation strategy to prevent over-fitting and to deal with the absence of knowledge about real sources. We thus apply a random permutation of the $T$ images of each spectral channel for each training sample $s$. This allows (i) to create artificially different datasets, and (ii) to break the temporal consistency of real sources. Synthetic sources are then  injected inside the temporally permuted data following the direct model (\ref{eq:image_formation_model}) to form the intermediate datasets $\lbrace \widebar{\V r} \in \mathbb{R}^{N \times T \times L} \rbrace_{s=1:S}$ such that:
\begin{equation}
	\widebar{\V r}_\ell^{[s]} = \M P^{[s]} \, \V f_\ell + \sum\limits_{p=1}^{P^{[s]}} \alpha_{p,\ell}^{[s]} \, \V h_\ell (\phi_p^{[s]})\,,
	\label{eq:injection_model}
\end{equation}
where operator $\M P$ performs the random temporal permutation. 
For each synthetic source $p$, we select randomly a synthetic SED $\V \alpha_p$ in $\mathbb{R}^L$ from a custom library of 10,000 sub-stellar spectra generated with ExoREM \cite{charnay2018self}, a physics-based model accounting for chemistry and for the presence of clouds in the exoplanet's atmosphere.  The number $P$ of injected sources is drawn uniformly in $\llbracket 1 ; 10\rrbracket$, since we expect few sources in the field of view. The initial source locations $\lbrace \phi_p \rbrace_{p=1:P}$ are drawn uniformly per angular separation, and their mean fluxes (integrated over the SED) are drawn uniformly in $\left[ 3\widehat{\sigma}_{\phi_p}^{\text{PACO}} ; 12\widehat{\sigma}_{\phi_p}^{\text{PACO}} \right]$, where $\widehat{\sigma}_{\phi_p}^{\text{PACO}}$ is the 1-sigma detection limit reached by \PACO at location $\phi_p$. We thus cover both sources that are detectable and unseen at the classical $5\sigma$ detection confidence using \PACO. 
As pre-processing is computationally demanding, we opt for a local update strategy. Prior to the injection of synthetic sources, the dataset $\V r$ is pre-processed to form $\widetilde{\V r}$, see Section \ref{subsec:pre_processing}. After injections, the set $\mathbb{S}^{[s]}$ of locations impacted by the signal of the $P^{[s]}$ sources is determined.
Outside $\mathbb{S}^{[s]}$, 
the pre-processed images are obtained from the temporal permutation of $\widetilde{\V r}$.
Inside $\mathbb{S}^{[s]}$, the statistics of the nuisance and the pre-processed images are updated given the contamination of the $P^{[s]}$ injected sources to form $\lbrace \widecheck{\V r} \in \mathbb{R}^{N \times T \times L} \rbrace_{s=1:S}$ such that:
\begin{equation}
	\widecheck{\V r}_{n,t,\ell}^{[s]} = \begin{cases}
  \M W_{n,t,\ell} \, \widebar{\V r}_{n,t,\ell}^{[s]}, & \text{for } n \in \mathbb{S}^{[s]}\, \cap\, \mathbb{P}\,, \\
	\M P^{[s]} \, \widetilde{\V r}_{n,t,\ell}, & \text{for } n \in \mathbb{P} - \mathbb{S}^{[s]}\, \cap\, \mathbb{P}\,.
\end{cases}
	\label{eq:preprocessing_local_update}
\end{equation}
Finally, the apparent motion (rotation and translation with respect to the star) of the sough objects are compensated to co-align spatially the signals of the injected sources along the temporal and spectral axis. It results $\lbrace \widebreve{\V r} \in \mathbb{R}^{N \times T \times L} \rbrace_{s=1:S}$.

\subsubsection{Loss function and metrics}
\label{subsubsec:loss_function}

\begin{spacing}{1}
The design of the training loss is driven by three criteria: (i) dealing with the strong class-imbalance, (ii) being computationally efficient, and (iii) representing a measure close to a detection accuracy score. We compared classical losses for pixel-wise classification, and we selected the Dice2 similarity score that was introduced first for biomedical image segmentation \cite{wang2020improved}. This loss satisfies the targeted criteria, in particular it penalizes equally errors in the source and background areas:
\end{spacing}
\begin{small}
\begin{equation}
	\mathcal{L}^{[s]}  = 1 - \underbrace{\frac{\sum\limits_{m} \V y_m^{[s]} \, \widehat{\V y}_m^{[s]} + \epsilon}{\sum\limits_{m} \V y_m^{[s]} + \widehat{\V y}_m^{[s]} + \epsilon}}_{\text{source error}}\, - \,\underbrace{\frac{\sum\limits_{m} (1-\V y_m^{[s]})(1-\widehat{\V y}_m^{[s]} + \epsilon)}{\sum\limits_{m} 2 - \V y_m^{[s]} - \widehat{\V y}_m^{[s]} + \epsilon}}_{\text{background error}}\,,
	\label{eq:loss}
\end{equation}
\end{small}
where $\lbrace  \V y^{[s]} ; \widehat{\V y}^{[s]} \rbrace$ is a set of ground truth and predicted detection maps, and $\epsilon$ is a small stability parameter whose setting has very limited impact on the detection performance. At validation time, we aim to obtain a model obeying a precision-recall trade-off. For a predicted detection map $\widehat{\V y}^{[s]}$ in $\left[ 0 ; 1 \right]^M$ thresholded at $\tau$ in $\left[0 ; 1\right]$, we derive the true positive rate (TPR) and the false discovery rate (FDR). 
From TPR and FDR, receiver operating curves (ROCs) are built by varying the detection threshold $\tau$. The area under ROCs (AUC) is used as an overall performance score. We observed experimentally that the loss (\ref{eq:loss}) is well anti-correlated to this accuracy metric.

\subsubsection{Architecture and implementation}
\label{subsubsec:model_architecture}

We select a U-Net with a ResNet18 as encoder backbone ($\simeq$11 millions of weights); an architecture widely used for image segmentation. Its implementation is based on the \texttt{SMP} package\footnote{\href{https://github.com/qubvel/segmentation\_models.pytorch}{https://github.com/qubvel/segmentation\_models.pytorch}}. The encoder and decoder parts have four blocks: each one containing a series of convolution, batch normalization,  activation (ReLU), and max pooling layers. The network ends with a sigmoid activation to produce a detection map $\widehat{\V y}$ in $\left[0 ; 1\right]^M$. At each training step $s$, the full stack $\widebreve{\V r}^{[s]}$ of temporo-spectral pre-processed images with synthetic sources is fed as input of the CNN. Pairs of samples $\lbrace \widebreve{\V r}^{[s]} ; \V y^{[s]} \rbrace$ are generated on the fly, and each realization $s$ is unique to prevent over-fitting. The notion of \textit{epoch} is used only as a way to evaluate regularly the performance and to schedule the learning rate. The number of samples $S$ per epoch is respectively set to 1,000 and to 100 at training and evaluation time. The batch size is equal to 1. The network is trained from scratch with AMSGrad. The parameters of the optimizer and of the scheduler have been fine tuned on a few datasets (weight decay: $10^{-5}$, learning rate: $10^{-3}$ with 10\% decrease every 10 epochs).

\section{Results}
\label{sec:results}

\subsection{Datasets Description and Tested Algorithms}
\label{subsec:datasets_algorithms}

The proposed algorithm is compared to the spectral version of baseline methods (cADI and PCA) as implemented in VIP \cite{VIP}, an open source package for HCI. 
 Details about the setting of these methods can be found in \cite{dPACO}. We also used the unsupervised \PACOII \cite{PACOII} algorithm to ground the benefits of the deep learning stage embedded in the proposed algorithm. To evaluate the benefits of a joint spectral processing, we also compare the proposed method against \PACO and \dPACO, that do not take benefits of the spectral diversity. The detection threshold is set to $\tau=5$ for cADI, PCA, \texttt{PACO (ASDI)}, which yields a SNR map as detection scores. 
 We set $\tau=0.5$ for $\dPACO$ and the proposed algorithm, that produce a pseudo-probability map. We selected three datasets obtained with the SPHERE-IFS instrument ($L=39$) by the observation of the following stars: $\beta$ Pictoris (observed in 2018-09-17, \textit{dataset 1}), $\beta$ Pictoris (2018-12-15, \textit{dataset 2}), and HR 8799 (2015-07-04, \textit{dataset 3}). We also selected the eleven datasets from the SPHERE-IRDIS instrument we considered in our recent work on \dPACO \cite{dPACO}. SPHERE-IRDIS produces ASDI datasets but in dual band only (i.e., $L=2$). In \cite{dPACO}, we processed the first spectral channel solely. In this paper, we revisit these datasets with the proposed method and we compare our results with \dPACO. We exemplify detailed results for three of these eleven datasets: HIP 88399 (2015-05-10, \textit{dataset 4}), HD 95086 (2021-03-11, \textit{dataset 5}), and HD 95086 (2015-05-05, \textit{dataset 6}).

\subsection{Qualitative and Quantitative Performance}
\label{subsec:quantitative_performance}

\begin{figure}[htbp]
	\begin{center}
		\includegraphics[width=0.48\textwidth]{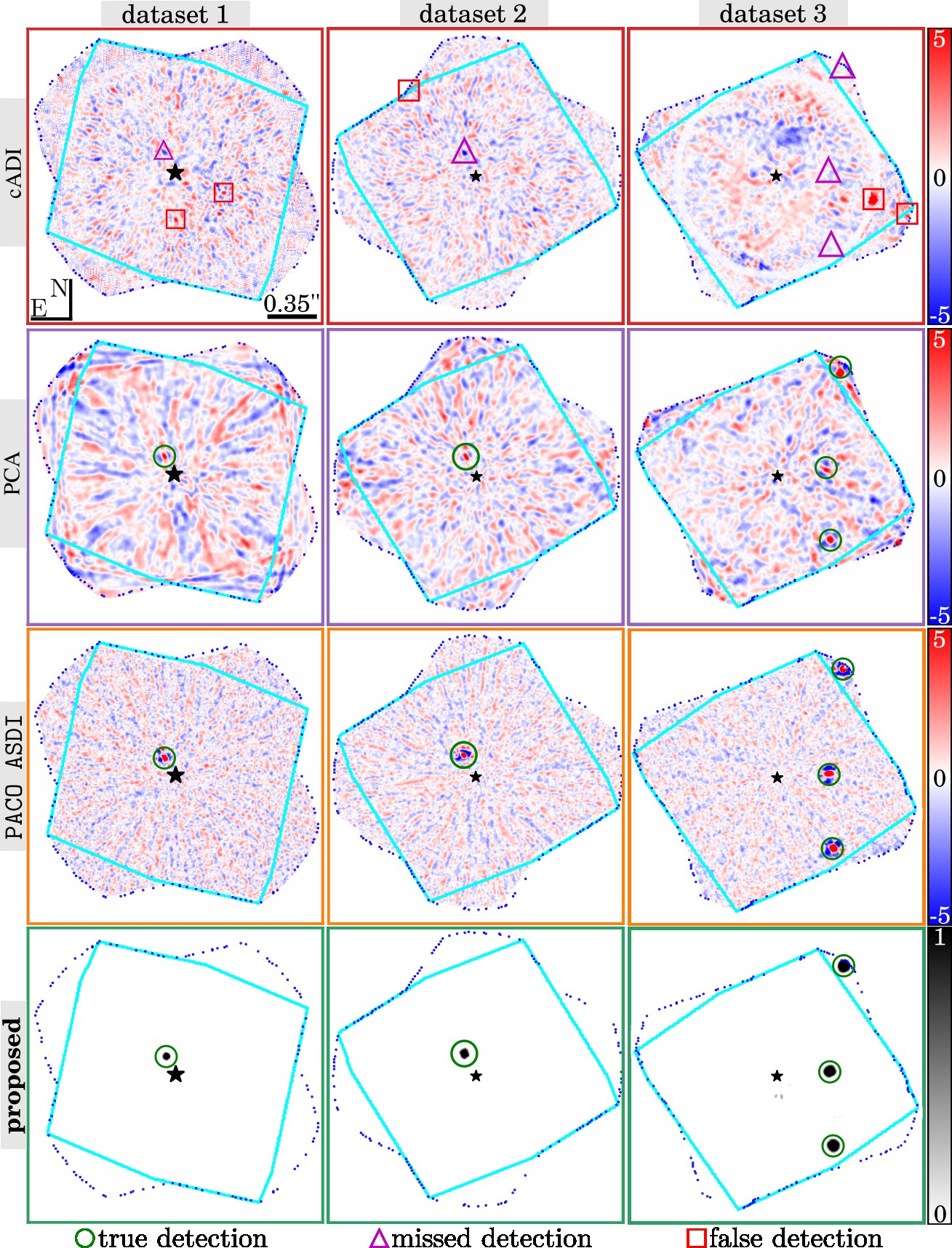}
		\caption{Detection maps obtained with the proposed approach on considered SPHERE-IFS datasets comparatively to three state-of-the-art methods in the field. Sources are classified as true, missed and false detections based on the prescribed detection threshold. Due to the apparent rotation of the field induced by ASDI, the detection can be performed on a larger field of view (dashed blue line) than the spatial extent of the sensor (light blue line).}
	\label{fig:detection_maps_ifs_real_data_vertical}
	\end{center}
\end{figure}

\begin{table}
	\caption{Mean AUC for ROCs giving the TPR as a function of the FDR for real and massively injected synthetic sources for SPHERE-IFS datasets. Best results are in bold font.
	}
	\label{tab:roc_mean_results_ifs}
	\centering
	\begin{footnotesize}
	\begin{tabular}{cccccc}
		\toprule
		source type & sep. ('') & cADI & PCA & \PACOII & \textbf{proposed}\\
		\midrule
		real & $\left[ 0.0 ; 1.0 \right]$ & 0.13 & 0.65 & 0.80 & \textbf{0.96}\\
		synthetic & $\left[ 0.0 ; 0.5 \right]$ & 0.24 & 0.36 & 0.68 & \textbf{0.76}\\
		synthetic & $\left[ 0.5 ; 1.0 \right]$ & 0.24 & 0.52 & 0.79 & \textbf{0.85}\\
		\bottomrule
	\end{tabular}
	\end{footnotesize}
\end{table} 

Figure \ref{fig:detection_maps_ifs_real_data_vertical} compares detection maps produced by the four multi-spectral algorithms we consider on the three datasets of SPHERE-IFS. The proposed algorithm, PCA and \PACOII are able to detect the five known exoplanets without any false alarm at the prescribed detection threshold. First line of Table \ref{tab:roc_mean_results_ifs} complements these results with AUC under ROCs built by varying the detection threshold. Even when the number of real sources is limited, these results emphasize that the proposed approach leads to a better trade-off between precision and recall than the other methods. We also resort to massive injection of simulated synthetic sources (considered in small batches) with various fluxes and positions to ground in more detail the performance of the proposed method. Second and third lines of Table \ref{tab:roc_mean_results_ifs} report the resulting AUC under ROCs for two regimes of angular separations (i.e., distance to the star and associated noise regimes). The proposed approach outperforms the other methods in both cases.

\begin{figure}[t!]
	\begin{center}
		\includegraphics[width=0.455\textwidth]{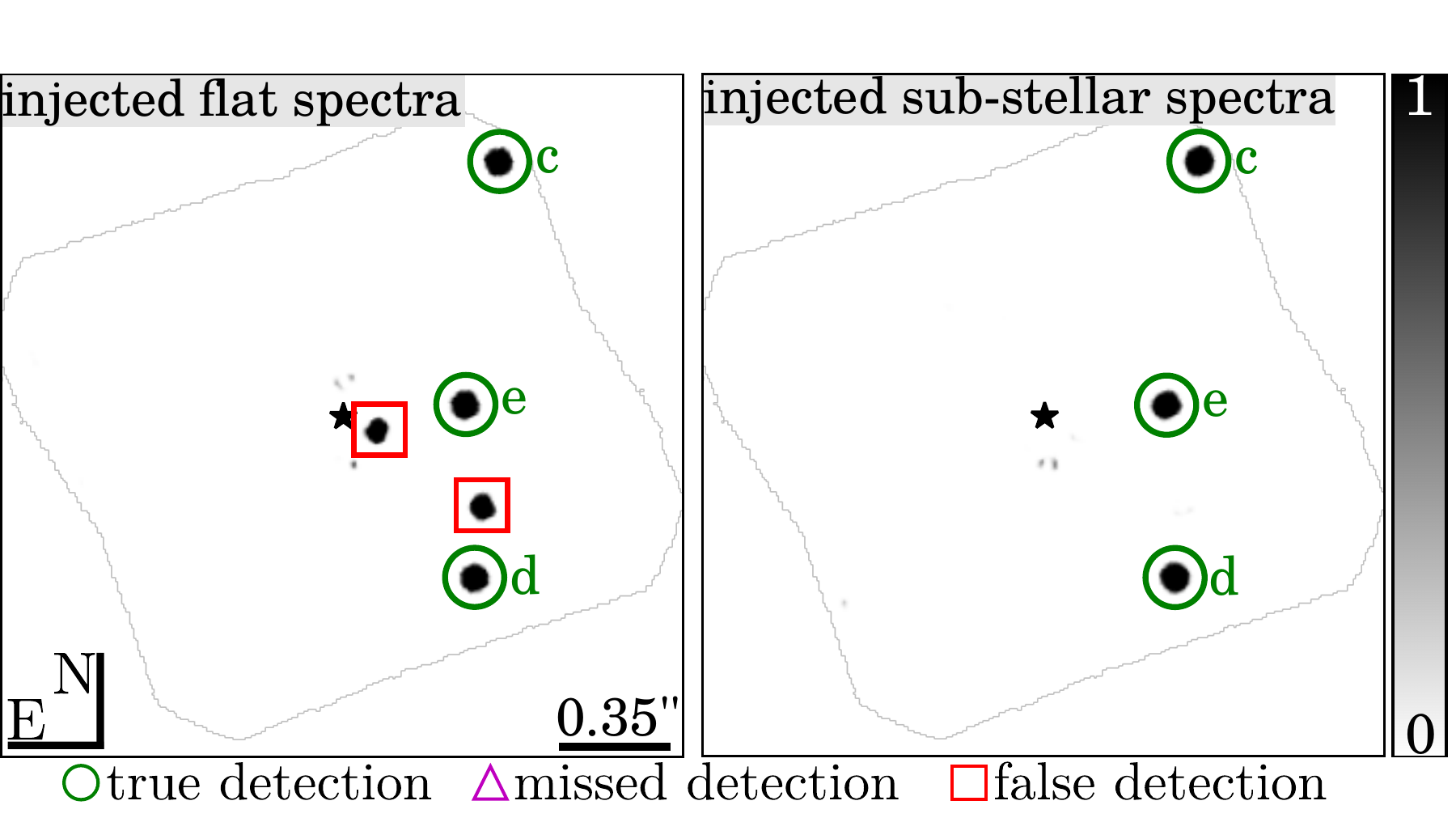}
		\caption{Detection maps obtained with the proposed algorithm on dataset 3. Left: injected training sources have a flat SED. Right: SED of injected sources is based on atmospheric models. The detection threshold is set to $\tau=0.5$, and sources are classified as true, missed and false detections. 
		}
	\label{fig:hr8799_ifs_detection_maps}
	\end{center}
\end{figure}

Figure \ref{fig:hr8799_ifs_detection_maps} completes this study by comparing detection maps produced by  the proposed algorithm using synthetic training sources having either a flat SED or realistic SEDs based on physical ExoREM models, see Section \ref{subsubsec:training_basis}. It emphasizes the importance of using reliable exoplanet SEDs since a flat SED (in contrast unit, i.e., same spectrum than the star) leads to additional detection peaks that we identified as false alarms based on the visual inspection of the detection maps and of the data themselves (i.e., the detection peaks are not static along the spectral axis or correspond to bright speckles).

\begin{figure}[t!]
	\begin{center}
		\includegraphics[width=0.48\textwidth]{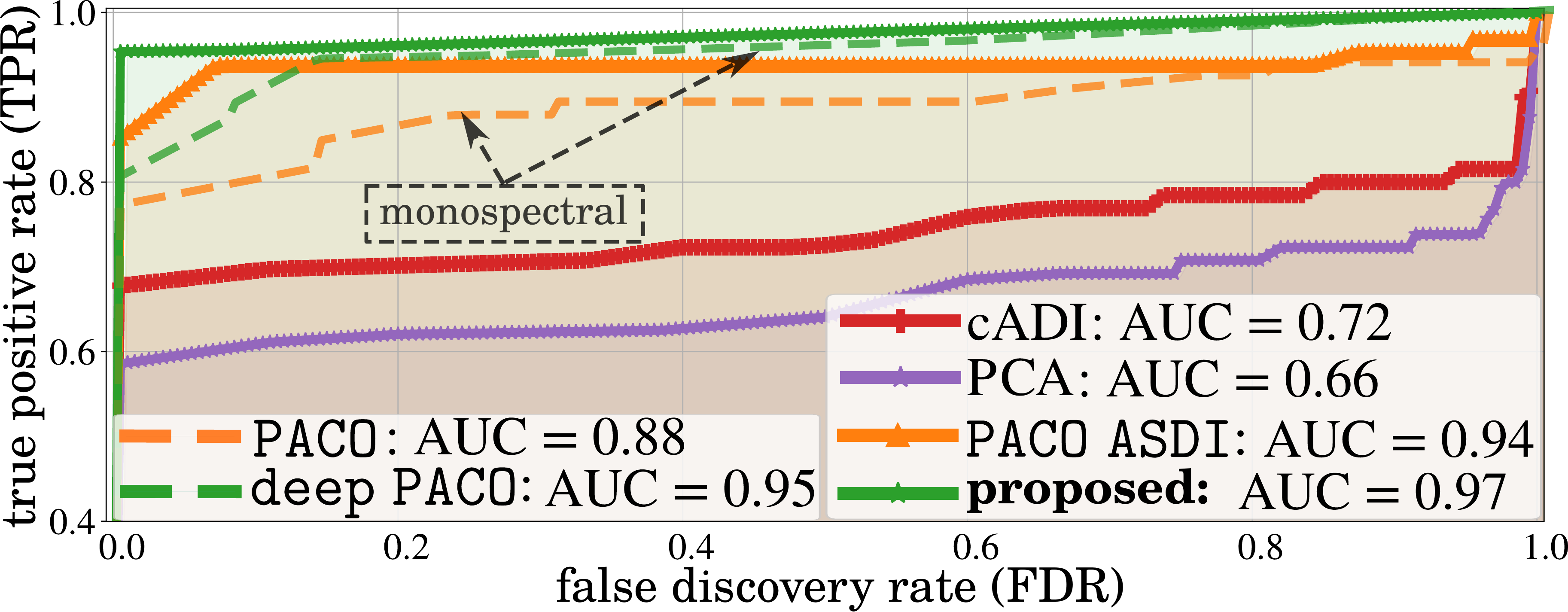}
		\caption{ROCs showing the TPR as a function of the FDR averaged over eleven SPHERE-IRDIS datasets. Straight lines are for a joint multi-spectral analysis (two channels), while dashed line stands for a monochromatic analysis.}
	\label{fig:irdis_asdi_roc_real_sources}
	\end{center}
\end{figure}

\begin{figure}[t!]
	\begin{center}
		\includegraphics[width=0.475\textwidth]{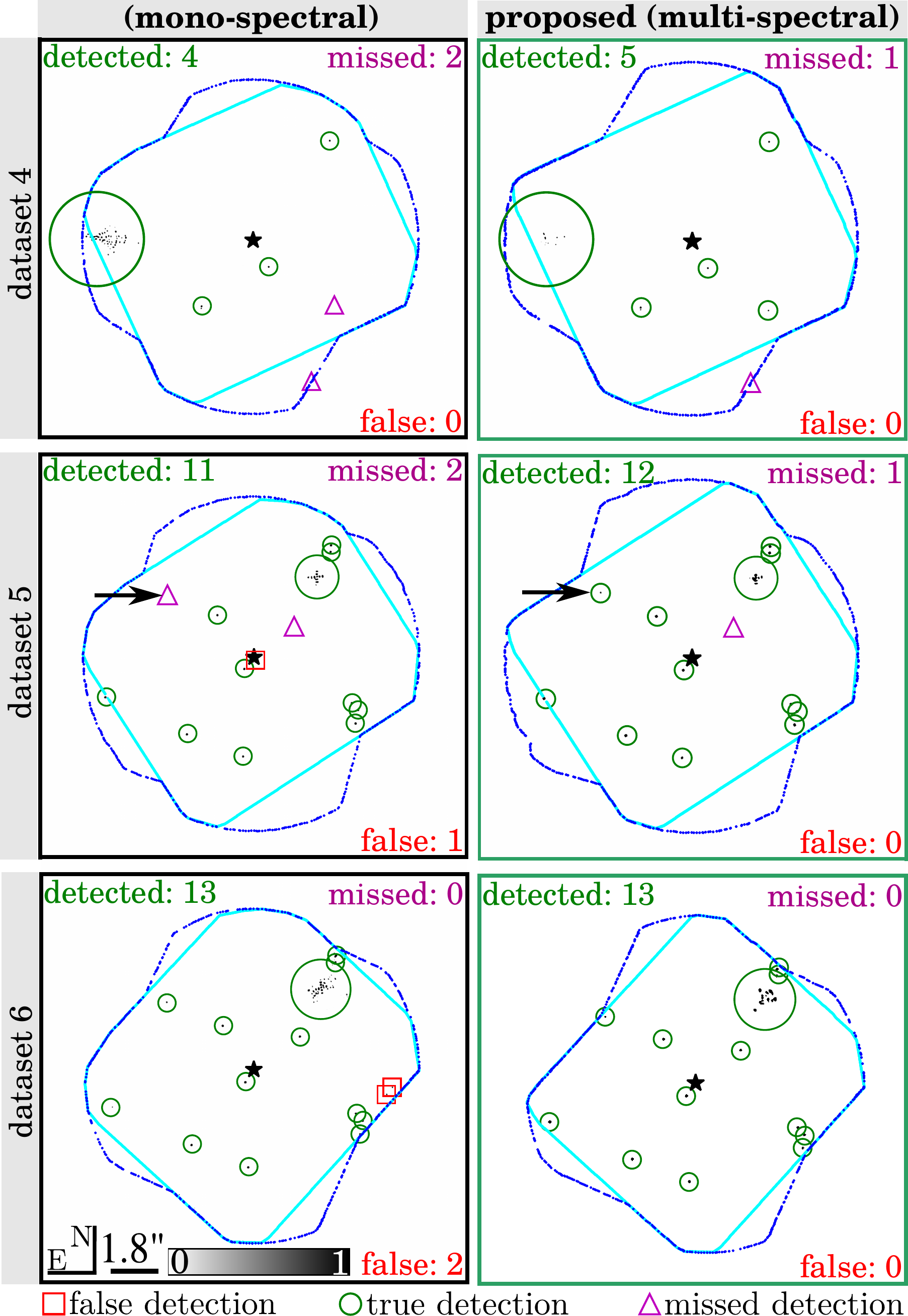}
		\caption{Comparison of SPHERE-IRDIS detection maps obtained by the proposed algorithm (right) and  its monochromatic version \cite{dPACO}.}
	\label{fig:irdis_adi_vs_asdi}
	\end{center}
\end{figure}

Finally, we briefly discuss the benefits of joint processing of multi-spectral data. Figure \ref{fig:irdis_asdi_roc_real_sources} gives ROCs averaged over the eleven SPHERE-IRDIS datasets (i.e., $L=2$) containing 59 off-axis real known point-like sources in total. These datasets were processed with the four considered multi-spectral algorithms. Our results are also compared to a monochromatic analysis (on the first spectral channel) with \PACO and \dPACO, as done in our recent work \cite{dPACO}. The detection maps from the datasets with the largest differences between mono- and multi-spectral processing are shown in Fig. \ref{fig:irdis_adi_vs_asdi}. These results illustrate the benefits of a joint multi-spectral analysis since it allows to avoid some false alarms, especially near the star where the nuisance component and its fluctuations are the strongest, as well as far from the star where border effects occur with mono-spectral processing (the two spectral channels having a slightly different field of view). Besides, the point-like source marked by a black arrow on dataset 5 corresponds to a new faint source (likely a background star seen in the projected field of view) that we have consistently identified on two datasets of the same star (including dataset 6) with \dPACO in \cite{dPACO}. It is now consistently detected on a third (worse) observation (dataset 5) by the proposed approach while it was not with its monochromatic version \cite{dPACO}.

\section{Conclusion}
\label{sec:conclusion}

We have proposed in this paper a new algorithm for exoplanet detection by the joint processing of spatio-temporal-spectral series of images. The algorithm includes a statistical modeling capturing most of the correlations of the data in a pre-processing step. A CNN is then trained to detect the residual signatures of injected synthetic sources. Our results show a better trade-off between precision and recall than state-of-the art methods. The joint exploitation of the spectral diversity improves the performance of our recent work based on a similar framework for spatio-temporal data. The control of the uncertainty and the construction of a model from multiple datasets jointly will be addressed in future work.



\vspace*{2ex}\noindent\begin{minipage}{\columnwidth}
	\footnotesize
\bibliographystyle{IEEEtran}
\bibliography{deep_paco_asdi_eusipco_2023}
\end{minipage}

\end{document}